\documentclass[sigconf,authorversion,nonacm]{acmart}
\AtBeginDocument{%
  \providecommand\BibTeX{{%
    \normalfont B\kern-0.5em{\scshape i\kern-0.25em b}\kern-0.8em\TeX}}}

\begin{document}

\title{A Language-Independent Analysis Platform for Source Code}

\author{Konrad Weiss}
\email{konrad.weiss@aisec.fraunhofer.de}
\orcid{0000-0002-1282-2162}
\affiliation{%
  \institution{Fraunhofer AISEC}
  \streetaddress{Lichtenbergstraße 11}
  \city{Garching near Munich}
  \state{Bavaria}
  \country{Germany}
  \postcode{85748}
}

\author{Christian Banse}
\email{christian.banse@aisec.fraunhofer.de}
\orcid{0000-0002-4874-0273}
\affiliation{%
  \institution{Fraunhofer AISEC}
  \streetaddress{Lichtenbergstraße 11}
  \city{Garching near Munich}
  \state{Bavaria}
  \country{Germany}
  \postcode{85748}
}

\renewcommand{\shortauthors}{Weiss and Banse}

\begin{abstract}
In this paper, we present the \textit{CPG} analysis platform, which enables the translation of source code into a
programming language-independent representation, based on a code property graph. This allows security experts and
developers to capture language level semantics for security analyses or identify patterns with respect to code
compliance. Through the use of fuzzy parsing, also incomplete or non-compilable code, written in different programming
languages, can be analyzed. The platform comprises an analysis library and interfaces to query, interact with or
visualize source code graphs. This set of CPG tools allows finding common weaknesses in heterogeneous software environments, independently of the
underlying programming language.
\end{abstract}




\maketitle

\section{Introduction}

Ensuring the correct behavior of software is crucial to avoid security issues stemming from incorrect implementations.
To ensure secure and compliant software even in the presence of ever-growing and more complex software systems, automated source code analysis tools are required.
More specifically, static analysis methods
are the key to consider all possible executions of a program and check whether they are running in conformance to
defined expectations. Large source code bases are difficult to analyze, as methods that exhaustively simulate program
states easily run into state explosion problems. Representing a finite source code base in a graph avoids such problems
while still representing all program executions. Query results on such graphs are complete but not sound representations of program execution.

Research into the graph-based analysis of source code has introduced the idea of a code property graph
(CPG)~\cite{yamaguchi2014modeling}. A CPG is a representation of source code in the form of a labeled
directed multi-graph. Each node and edge is assigned a set of key-value pairs, named properties. Nodes
represent syntactic elements of programming language, whereas edges capture the relations between them. These
relations can be single edges, e.g. edges from call expression to potential targets of the invocation, or connected
sub-graphs that form complex structures, e.g. control-flow and data-flow edges. 

In this paper, we present a language-independent analysis platform for source code based on an adaption of such a graph. With this analysis framework, we can run different analysis such as data flow analysis or type state analysis on a graph representation of source code. Our contributions are
\begin{itemize}
	\item A language-independent graph representation and query capability of source code
	\item An architecture of a source code analysis platform based on this graph
	\item An extensible open-source implementation\footnote{\url{https://github.com/Fraunhofer-AISEC/cpg}} of the platform to analyze Java, C/C++, Go, Python, TypeScript and LLVM-IR
\end{itemize}

\section{Design Goals}

With our platform, we aim to support security experts during an audit of source code and developers to perform an automated analysis e.g. in their CI/CD pipeline. These scenarios drive the following goals, that we aim to fulfill in the design
and implementation of our platform:

\begin{itemize}
  \item \textbf{G1: Allow analysis of incomplete code} to support its usage during the early development lifecycle and auditing where only parts of the code are available.
  \item \textbf{G2: Create a language-independent representation} to allow for definitions of rules for multiple languages
  and analysis of language-heterogeneous systems, such as the Cloud.
  \item \textbf{G3: (Semi-)automated use} to enable application in auditing as well as development environments.
  \item \textbf{G4: Model language level semantics} to increase precision by allowing differentiation according to nuanced
  language semantics.
\end{itemize}

\section{Graph Structure}\label{sec:codepropgraph}

In this section, we describe the structure of our CPG, which is used to represent source code of
different programming languages. This structure aims to provide a superset of language features found in most object-oriented languages, such as C++, Java or Go. This includes structural elements, such as functions, methods,
classes as well as expressions/statements, e.g., calls, operators, literals or conditions.

\subsection{Abstract Syntax Tree}\label{sec:astGraph}

An abstract syntax tree (AST) is a tree structure representing the syntactic elements of source code in its nodes. The root
the tree is a \textbf{TranslationUnitDeclaration} node, which represents the code contained in one file. The complete graph then comprises the set of all trees. Different types of AST child nodes exist in the tree, each representing different semantics within a program:

\begin{itemize}
  \item \textit{Structural Entities} represent entities that give the code its structure. Usually, they can be used
   as entities or instantiated as such. Examples are nodes representing namespaces, classes or structs.
  \item \textit{Value Declarations} are identifiers that contain or return values and therefore are used to model
  local variables, parameters, functions, methods and constructors.
  \item \textit{Nodes of Program Execution} model statements and expressions. Statements are syntactic units which are executed in
sequence and thus form the program logic of imperative programs. In contrast, expressions are syntactic units which
evaluate to a value and are of a specific value type. Expressions can be nested and can be a child of a statement.
\end{itemize}
The individual node types are realized through sub-classing and therefore model a semantically rich hierarchy,
represented by Java classes\footnote{See \url{https://fraunhofer-aisec.github.io/cpg/} for a complete model of the graph
in the \texttt{de.fraunhofer.aisec.cpg.graph} package} in our reference implementation and by node labels in the graph
which contributes to \textbf{G4}. For example, a \textbf{MemberCallExpression}, representing a call to a class member,
inherits from \textbf{CallExpression}, which in turn is derived from \textbf{Expression}.

The edges of the AST represent the directed parent-child relation between syntactic elements. The edge also stores the
information which relation the child holds to the parent, e.g. \textbf{LHS} for the left-hand side of an expression.
Indices are stored on edges wherever the syntactic order is relevant for program execution, as graphs generally do not
support the concept of order between nodes.

The following semantics of program execution, such as control and data flow, are then modeled as edges between the AST nodes.

\subsection{Control Flow and Evaluation Order}
\label{subsec:eog} 

The CPG adapts the concept of a control-flow graph (CFG) into an evaluation order graph (EOG), that interconnects
statements and expressions in the order that they are evaluated, to represent control flow on a finer-grained level.
This is necessary to correctly capture side effects that come from the order of execution inside an expression, e.g.
\texttt{a() + b()} or \texttt{a > 1 ? a : 0}. To build the EOG of an expression \texttt{a + 5}, first
\texttt{a}, then \texttt{5} and lastly their common parent node \texttt{a + 5} is connected, similar to a post-order
traversal of these nodes. This model of evaluation order follows the notion a compiler would follow for expression
evaluation, and is extended to non-expression nodes that have an execution order. The evaluation order is from left to right with few exceptions:
\begin{itemize}
	\item Constructs that explicitly change the control flow of a program, such as \texttt{if} or other conditional expressions.
	\item Nodes representing code that is not executed in the order of code appearance, e.g. the body of a for-loop being
	executed before its iteration expression, but after the initializer.
\end{itemize}

The root node of branching nodes is connected after the branching expression, e.g. condition or selector, and before the
branching targets to allow algorithms that have traversed the branching expression to get information on the semantic
root before having to handle the branch. EOG edges at such branching positions save additional information on the result
of the branching expression that leads to the branch's execution, e.g. \texttt{true} or \texttt{false} for conditions.

\subsection{Data Flow}\label{subsec:dfg}

Operations or entities that handle data are represented in a data-flow graph (DFG) within the CPG. To model data flows through the program, the following \textit{DFG} edges connect nodes:
 \begin{itemize}
   \item A child contains an edge to its parent if the parent's value depends on the child's value.
   \item A referenced variable is the start or end of an DFG edge if the variable is written to or read from.
   \item Edges are drawn from argument expressions of a call to the respective parameters of the call target.
   \item \textit{either} an edge is drawn from the write reference to the referenced declaration
   \item \textit{or} an edge is drawn from the last valid writing expression to the \textit{point-of-use} if the control-flow-sensitive data flow analysis was enabled.
 \end{itemize}
 The resulting data-flow sub-graph is control-flow sensitive but not context-sensitive.

\subsection{Additional program semantics}\label{subsec:other}

The CPG contains additional nodes and edges to model program semantics that are necessary for program analysis:

\begin{itemize}
  \item The type system of a program and used language is modeled by adding its complex hierarchical structure as
  nodes and edges forming a type sub-graph, that contributes to \textbf{G4}.
  \item \textit{REFERS\_TO} edges are drawn between references and the declaration they target.
  \item \textit{INVOKES} edges are drawn between calls and identified call targets in a best-effort approach to give an
  inter-procedural extension to the intra-procedural EOG.
\end{itemize}

\begin{figure*}
  \begin{center}
    \includegraphics[width=.8\linewidth]{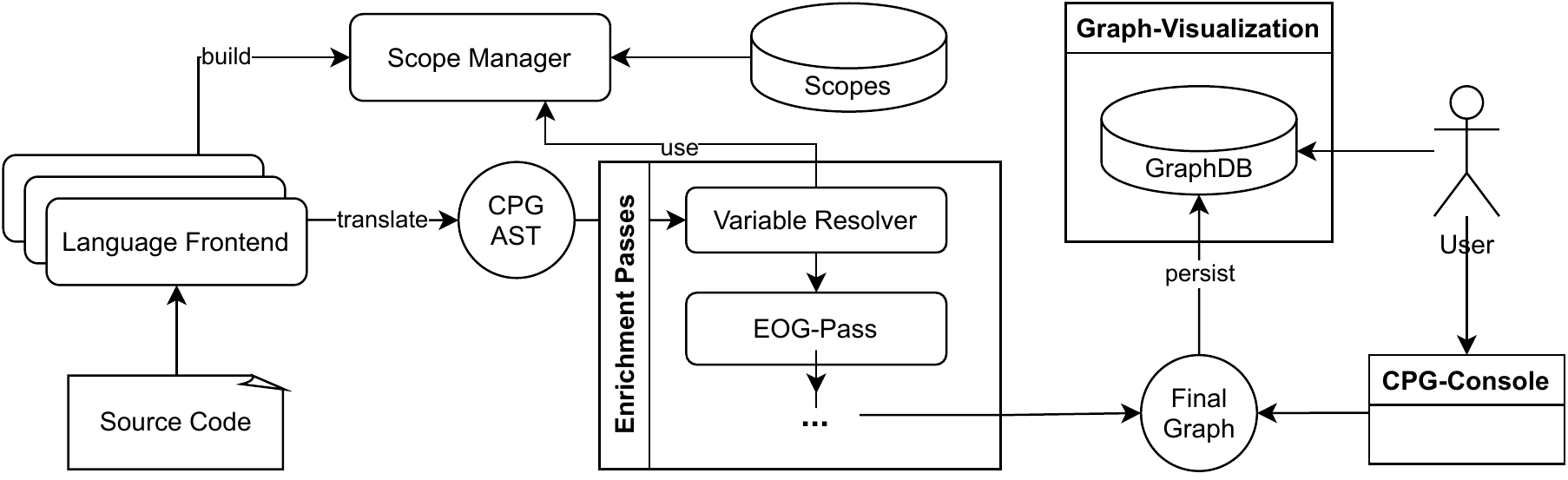}
  \end{center}

  \caption{CPG analysis platform showing the workflow when analyzing source code}

  \label{fig:architecture}
\end{figure*}

\section{Architecture and Implementation}\label{sec:architecture} The CPG project is composed of several tools
built around a library for iterative graph construction of source code in form of a Java/Kotlin-based open source
implementation\footnote{\url{https://github.com/Fraunhofer-AISEC/cpg}}.
Figure~\ref{fig:architecture} depicts the
workflow of translating source code into the graph representation as well as its visualization and analysis by a user.
The following sections will elaborate on the individual components.

\subsection{Analysis Library}
The CPG library comprises several components, which are used to configure the analysis and translation of heterogeneous
code into an in-memory graph structure. Java, C and C++ are the main languages currently supported by the platform. Additionally,
experimental frontends for Python, Golang, TypeScript and LLVM-IR exist. 

\paragraph{\textbf{Language Frontends}}

Source code files are dispatched to a language-specific frontend by their file extension. The frontend can then either
use a parsing library accessible from the JVM or use the Java Native Interface (JNI) to use language-native techniques
to retrieve the AST for a particular language\footnote{We make use of JNI in our experimental frontends for Go and
Python}. A language frontend is expected to perform the following tasks:
\begin{itemize}
	\item Language dependent AST children are translated into the language independent CPG AST structure to enable
	\textbf{G2}, as described in Section~\ref{sec:astGraph}.
	\item Entities that are implicitly present but not located in the source code are added explicitly to the graph, e.g.
    implicit \texttt{this} fields and missing \texttt{return} statements.
	\item Identifiers are collected in a scope tree to later resolve access to names in a fuzzy manner.
\end{itemize}

\paragraph{\textbf{Scope Manager}}
In most programming languages, the declaration of a name is not globally valid, but its validity is restricted to an
area associated with a language construct, such as a class or a function. This area of validity is called a
scope. The scope manager tracks the currently active scope stack while the frontend traverses AST nodes. 
This allows for tracking and resolving declarations by absolute or relative name and
managing control flow jumps that are bound to the scope of an enclosing language feature, e.g. loops, try-statements.
After building the scopes in the language frontends, the scope manager holds several scope trees that allow random
access in future passes.

\paragraph{\textbf{Passes}}

As mentioned before, the frontend produces partially connected AST trees. Afterwards, passes are used to enrich the CPG
by implicit execution information and program semantics such as usage-references,
data-flows, and evaluation order. These semantics are
built between the language independent AST nodes and are themselves language independent contributing to \textbf{G2}.
However, they still allow for language-specific customization. Passes depend on the prior execution of other passes when
needing their semantics in the graph. Lastly, passes also support inference. Nodes are added and marked as
\texttt{IMPLICIT}, for entities that are not directly visible in source code, or \texttt{INFERRED}, when part of missing
source code. This allows for fuzzy parsing and analysis of incomplete code and works towards \textbf{G1} by adding
missing declarations of entities due to missing dependencies or code components.

The library's functionality can be extended by any dependent applications through registration of newly implemented language frontends and passes or adapting existing components through subclassing and overwriting.

\subsection{Persistence and Visualization}

The CPG tool contains a persistence component that stores the in-memory graph into a Neo4J graph database. This
allows manual exploration through interaction in a visual interface, as well as running
arbitrary analysis queries written in graph-query languages.

\subsection{CLI Console}

\begin{table*}
  \caption{Runtime and coverage evaluation of 100 Repositories per Language.}
  \centering
  \footnotesize
  \begin{tabular}{|c|c|c|c|c|c|c|c|c|c|} 
   \hline
   \textbf{Lang.} & \textbf{Repos[\#]} & \textbf{Total ET[s]} & \textbf{ET Passes[\%]} & \textbf{Total SLoC[\#]} &
   \textbf{ET / SLoC[ms]} & \textbf{Avg. SLoC[\#]} & \textbf{Cov.[\%]} & \textbf{Uncov.[\%]} &
   \textbf{Partial[\%]} \\
   \hline
   Java & 97 & 1042.10 & 37.5 & 211,541 & 4.92 & 2180 & 99.16 & 0.7 & 0.12 \\
   \hline
   C++ & 88 & 687.98 & 46.0 & 148,036 & 4.65 & 1682 & 96.10 & 3.8 & 0.05 \\
   \hline
  \end{tabular}
  \hspace{1em}
  \label{tab:evaluation}
\end{table*}

Our platform offers a simple shell based on the interactive Kotlin interpreter
\textit{ki}\footnote{\url{https://github.com/Kotlin/kotlin-interactive-shell}} which supports
semi-automated and manual analysis by auditors, as well as automatic runs of security checks for development contexts, therefore realizing our goal \textbf{G3}.
This is achieved by providing functionality for the traversal and inspection of the in-memory graph, an extendable set
of built-in commands for graph-interaction as well as a collection of security related analysis examples, such as a
null-pointer detection.


\section{Evaluation}

In this section, we assess the viability of the CPG platform in terms of runtime performance and code coverage. 
The evaluation is conducted over a set of 100 Java and 100 C++ open source repositories that we arbitrarily selected on GitHub. 
Because analysis passes are executed after all files are translated and use
algorithms whose execution times are not necessarily linear to the lines of code, we compute the execution times based
on the overall repositories' SLoC and not on individual files. The collection of the repositories is not part of
the measurements. The evaluation was performed on a virtual machine with a 3.3 GHz vCPU, 63 GiB
of Memory and 388 GB of disk space running Ubuntu 20.04.

\paragraph{\textbf{Execution Time (ET)}}

We measure the execution time (\textit{ET}) in \textit{seconds / SLoC} to give an impression of the expected
runtime based on actual source code lines. By measuring a metric relative to the size of the repository, we aim to
counter the effect that the repositories in the evaluation data set were of different sizes.

\paragraph{\textbf{Code Coverage}} 

Similar to \textit{code coverage} in unit testing frameworks, we built a code coverage metric that shows how many of the
original language constructs, based on the AST, were successfully represented in our final graph. The coverage is based
on \textit{SLoC}, with a line associated with an AST node being counted as:
\begin{itemize}
  \item \textit{uncovered}, if no handler was implemented for the AST node
  \item \textit{covered} if a leaf-node was properly handled or if the children of a non-leaf node were finished processing
  and the line is not in any of the children's \textit{uncovered} or \textit{partial} set.
  \item \textit{partial}, if the line is contained in one child's \textit{uncovered} set and another child's \textit{covered} set.
\end{itemize}

The algorithm to compute coverage sets in \textit{SLoC} is defined recursively and shows inaccuracies in the
representation. All remaining source lines, i.e. those not contained in any set so far, are added to \textit{uncovered},
as these lines were not visited.
Note that in contrast to AST children that cannot be handled, AST children that are not forwarded to the handlers cannot be counted and their respective code is added to \textit{covered}. For this reason, the results have to be considered an upper bound to the coverage.

\paragraph{\textbf{Discussion}}

Table~\ref{tab:evaluation} shows the results of the analysis. 3 Java and 12 C++ repositories did exceed the maximum
analysis time of 5 minutes and ran into a timeout. We did not consider them in the final analysis, which may limit
representability of larger code repositories. Based on the remaining repositories, 211,541 and 148,036 SLoC were
analyzed in 1042 and 687 seconds, respectively, which suggests that the platform can be used to analyze small to medium
code repositories in acceptable time. The runtime per SLoC can be used to estimate the analysis runtime of larger
repositories for the specific language. However, non-linear increases in runtime could appear for larger code
repositories. 

While the analysis of C++ Code was faster per SLoC, the coverage metrics show that this may be due to slightly less code
coverage. The execution of graph enriching passes took up less than half of the total time, 37 \% for Java and 46 \% for
C++. This is not surprising as most enriched semantics are intra-procedural. The coverage metric was previously
mentioned to be imprecise and represents an upper bound to the handled source code elements. In conjunction with a lower
bound metric, this upper bound would allow assessing the effectiveness of the current implementation. Upper bounds of
99\% and 96\% are not surprising as the implementation of analysis tools puts priority on frequently used language
features.

\section{Related Work}

Existing tools and techniques that build CPG analysis differ in their level of abstraction and their support for
programming languages. For example, \textit{Joern}~\cite{Joern} is a security analysis platform for several languages,
such as C/C++, Java or JavaScript. \textit{Plume}~\cite{Plume} and \textit{Graft}~\cite{Graft} represent tools that
translate Java byte-code into a graph structure. Next to CPG-related tools, \textit{CodeQL} (formerly known as
Semmle)~\cite{codeql} is a query language and engine for semantic code analysis with extensions that are specific to the
supported programming languages.

Our platform differs from other tools with respect to the degree of extensibility and language abstraction. Providing an
extensible platform is one of the declared goals of the CPG tool. This is achieved by providing a well-defined API to
applications that allows to register new language frontends (to add support for additional languages) or passes that add
additional semantics. When adding support for additional languages, developers only need to focus on translating an AST
provided by a language parser into the generic CPG AST nodes (see Section~\ref{sec:astGraph}). All other steps, such as
call resolving, control- and data-flow construction will be executed by the existing language-independent passes. This provides reusability of implementations similar to \textit{Joern} but differing both from \textit{CodeQL}.

In its level of language abstraction, the CPG tool meets a balance between \textit{Joern} and \textit{CodeQL}. We keep
more language-specific semantics by modeling a more differentiated set of AST nodes and type hierarchy than
\textit{Joern}, while keeping resulting queries language independent. \textit{CodeQL} in contrast, uses a loose system
of generic AST type-interfaces and highly language-specific implementations that do not allow for language-independent
queries.

\section{Conclusion}

In this paper, we present the \textit{CPG tool}, a platform of tools to analyze source code written in different
programming languages using a uniform graph representation. We show, using an evaluation of 200 source code
repositories, that the CPG is suitable to analyze small to medium-size repositories, independently of the programming
language. It offers coverage of the most common language constructs, especially for the C/C++ and Java languages. The
evaluation of our translation execution times, shows that suitable scenarios include security audits or checks during
CI/CD runs. However, further improvements are necessary to allow real-time code analysis in the early development cycle.
Future work, therefore, includes the parallelization of code translation as well as the incremental construction of
graphs and program semantics.

\bibliographystyle{ACM-Reference-Format}
\bibliography{ase-bibliography}


\begin{thebibliography}{5}


\ifx \showCODEN    \undefined \def \showCODEN     #1{\unskip}     \fi
\ifx \showDOI      \undefined \def \showDOI       #1{#1}\fi
\ifx \showISBNx    \undefined \def \showISBNx     #1{\unskip}     \fi
\ifx \showISBNxiii \undefined \def \showISBNxiii  #1{\unskip}     \fi
\ifx \showISSN     \undefined \def \showISSN      #1{\unskip}     \fi
\ifx \showLCCN     \undefined \def \showLCCN      #1{\unskip}     \fi
\ifx \shownote     \undefined \def \shownote      #1{#1}          \fi
\ifx \showarticletitle \undefined \def \showarticletitle #1{#1}   \fi
\ifx \showURL      \undefined \def \showURL       {\relax}        \fi
\providecommand\bibfield[2]{#2}
\providecommand\bibinfo[2]{#2}
\providecommand\natexlab[1]{#1}
\providecommand\showeprint[2][]{arXiv:#2}

\bibitem[\protect\citeauthoryear{Effendi et~al\mbox{.}}{Effendi
  et~al\mbox{.}}{2021}]%
        {Plume}
\bibfield{author}{\bibinfo{person}{David~B. Effendi} {et~al\mbox{.}}}
  \bibinfo{year}{2021}\natexlab{}.
\newblock \bibinfo{title}{Plume}.
\newblock \bibinfo{howpublished}{\url{https://github.com/plume-oss/plume}}.
\newblock


\bibitem[\protect\citeauthoryear{GitHub}{GitHub}{2021}]%
        {codeql}
\bibfield{author}{\bibinfo{person}{Inc. GitHub}.}
  \bibinfo{year}{2021}\natexlab{}.
\newblock \bibinfo{title}{CodeQL}.
\newblock \bibinfo{howpublished}{\url{https://codeql.github.com/}}.
\newblock


\bibitem[\protect\citeauthoryear{Keirsgieter}{Keirsgieter}{2021}]%
        {Graft}
\bibfield{author}{\bibinfo{person}{Wim Keirsgieter}.}
  \bibinfo{year}{2021}\natexlab{}.
\newblock \bibinfo{title}{Graft}.
\newblock \bibinfo{howpublished}{\url{https://github.com/wimkeir/graft}}.
\newblock


\bibitem[\protect\citeauthoryear{Yamaguchi et~al\mbox{.}}{Yamaguchi
  et~al\mbox{.}}{2021}]%
        {Joern}
\bibfield{author}{\bibinfo{person}{Fabian Yamaguchi} {et~al\mbox{.}}}
  \bibinfo{year}{2021}\natexlab{}.
\newblock \bibinfo{title}{Code Property Graph - Specification and Tooling}.
\newblock
  \bibinfo{howpublished}{\url{https://github.com/ShiftLeftSecurity/codepropertygraph}}.
\newblock


\bibitem[\protect\citeauthoryear{Yamaguchi, Golde, Arp, and Rieck}{Yamaguchi
  et~al\mbox{.}}{2014}]%
        {yamaguchi2014modeling}
\bibfield{author}{\bibinfo{person}{Fabian Yamaguchi}, \bibinfo{person}{Nico
  Golde}, \bibinfo{person}{Daniel Arp}, {and} \bibinfo{person}{Konrad Rieck}.}
  \bibinfo{year}{2014}\natexlab{}.
\newblock \showarticletitle{Modeling and discovering vulnerabilities with code
  property graphs}. In \bibinfo{booktitle}{\emph{2014 IEEE Symposium on
  Security and Privacy}}. IEEE, \bibinfo{pages}{590--604}.
\newblock


\end{thebibliography}

\end{document}